\documentclass{elsarticle}
\usepackage[utf8]{inputenc}
\usepackage{graphicx}
\usepackage{hyperref}
\usepackage{url}
\newcommand{\COO}{CO$_2$ }
\usepackage{subcaption}
\usepackage{amsmath}
\usepackage{amsfonts}
\usepackage{xcolor}
\usepackage{siunitx}

\begin{document}

\begin{frontmatter}

\title{Optimised operation of low-emission offshore oil and gas platform integrated energy systems}
\author{Harald G Svendsen}\ead{harald.svendsen@sintef.no}
\address{SINTEF Energy Research, Sem Sælands vei 11, Trondheim, Norway}

\begin{abstract}
This paper considers the operation of offshore oil and gas platform energy systems with energy supply from wind turbines to reduce local \COO emissions.

A new integrated energy system model for operational planning and simulation has been developed and implemented in an open-source software tool (Oogeso). This model and tool is first presented, and then applied on a relevant North Sea case with different energy supply alternatives to quantify and compare emission reductions and other key indicators.
\end{abstract}

\begin{keyword}
integrated energy system \sep optimisation \sep oil and gas \sep operational planning
\end{keyword}

\end{frontmatter}

\section{Introduction}

Offshore oil and gas platforms \cite{devold2013} are typically isolated systems with local energy supply by natural gas turbines.
In the past, offshore energy system operation and power management systems have relied on the high controllability of gas turbines and an excess of low-cost natural gas fuel, with few incentives to consider alternatives. This is presently changing with the need for \COO emission reductions due to policy targets, legal limits or high carbon emission taxes.
Local emissions from oil and gas fields come mainly from the combustion of natural gas to supply local energy demand, and to eliminate these emissions, fossil-fuel based gas turbines must be replaced by clean alternatives, such as power from offshore wind, power via cable from shore, shift to hydrogen-based gas turbine fuels, or fuel cells. 
With new energy supply alternatives, new operating strategies are required in order to best utilise the available resources within the given constraints.


Better energy efficiency is one way to reduce \COO emissions. A potential for energy savings and emission reductions up to about  15--20\% has been indicated in the literature \cite{NGUYEN2016325}. Improved efficiency, especially related to production manifolds and gas compression systems, is important \cite{VOLDSUND201445}. However, energy efficiency alone can never bring emissions anywhere close to zero.
Gas turbines with carbon capture and storage may be possible \cite{ROUSSANALY2019478,NORD20176650}, but is unlikely to ever become a practical and economical option for the small gas turbines used on oil and gas platforms. 

To achieve large \COO emission reductions, it seems clear that a different energy supply is needed. Electric power via cables from shore is an economical option for near-shore installations that reduces emissions by replacing inefficient offshore gas turbines by better alternatives onshore \cite{riboldi2019}. However, cables are costly, and this approach adds strain on the onshore system in terms of increased power demand and need for grid transmission capacity. 
Gas turbines running on hydrogen fuels is another alternative being considered \cite{AESOY2020269}, but is not available yet.

Electrification is important, but not the whole solution. A significant portion of the energy-demanding equipment on existing offshore installations is mechanically supplied via direct drive gas turbines \cite{kurz2005}, and switching to electric drives is not straightforward. Moreover, as heat is normally supplied via gas turbine waste heat recovery, a full replacement of the gas turbines requires new systems for heat supply.

Offshore wind power is an obvious alternative that is already available at commercial scale, and the oil and gas industry has already shown interest in offshore wind power development \cite{MAKITIE2019269}. 
Presently, the interest is very high, exemplified by the Hywind Tampen development where an 88 MW wind farm will supply power to the Snorre and Gullfaks oil and gas fields \cite{hywindtampen} from 2022.

A study  of the potential for emission reductions through improved operational strategies with hybrid wind and gas power supply to an oil and gas field  has highlighted that the appropriate sizing of the  wind power plant is crucial for the cost-effectiveness of such a hybrid system \cite{KORPAS201218,he2010}. The study considered wind energy share up to \SI{40}{\%}.
Increasing wind power capacity further gives oversupply when the wind is strong, and still lack of power when the wind is not blowing. This is in line with the actual development at Hywind Tampen, where the wind farm capacity does not exceed the power demand, relying on gas turbines to provide balancing and the remaining energy.
A study of an oil production platform in the Pacific found an optimal wind penetration level of 57\% with resulting emission reduction of 40\%, using a mixed-integer linear optimisation approach \cite{ZHANG2021126225}. However, it is unclear how wind power variability is accounted for and whether power system stability issues are considered in this study.

Electrical stability of oil and gas installations with wind power supply has been studied both for isolated systems \cite{svendsen6019309,ARDAL2012229,4523677}
and for systems with cable link to shore \cite{MARVIK2013558,kolstad6608004}, with conclusions indicating that such systems can perform well with appropriate energy management and electrical system controls \cite{Xie2018}.
In systems with large energy supply and demand variability, energy storage may be needed both for energy  balancing and for electrical stability  \cite{10.3389/fenrg.2021.649200}.
A challenge with high shares of converter interfaced load and generation is that it gives volatile electrical systems with low inertia, requiring new types of control \cite{mota2021a,mota2021b}.
Alternatively, or in addition, flexibility in loads, such as water injection pumps with variable speed drives, can alleviate some of the challenges related to the variability of wind power supply, even stability issues due to short-term wind-induced power fluctuations \cite{sanches2017}.


With a shift away from gas turbines, new energy system operational strategies considering all sources of variability and flexibility are needed, both for maintaining the security of supply and for keeping operating costs at a minimum. 
This is especially important for electrically isolated systems without cable to shore. 
To assist the development and assessment of such new energy system designs and operating strategies for offshore energy systems, improved analysis and optimisation models are needed. 
This paper contributes in this direction with a focus on the integration of wind power and batteries to reduce emissions.

Integrated multi-energy modelling and optimisation of offshore oil and gas energy systems is not new, but previous studies \cite{nguyen2012,zhang2019} are mainly concerned with traditional systems with gas turbines, considering \COO emission reductions, but not to the degree needed to meet industry and societal targets for 2030 and beyond. To become carbon neutral, design and operation of platform energy systems will need to change dramatically.
Oil and gas platforms in the North Sea are part of an energy infrastructure that will undoubtedly undergo a big transition in the coming years \cite{MCKENNA2021100067}, where offshore energy hubs may become key building blocks \cite{zhang2021modelling}.

\begin{figure}
    \centering
    \includegraphics[scale=0.5]{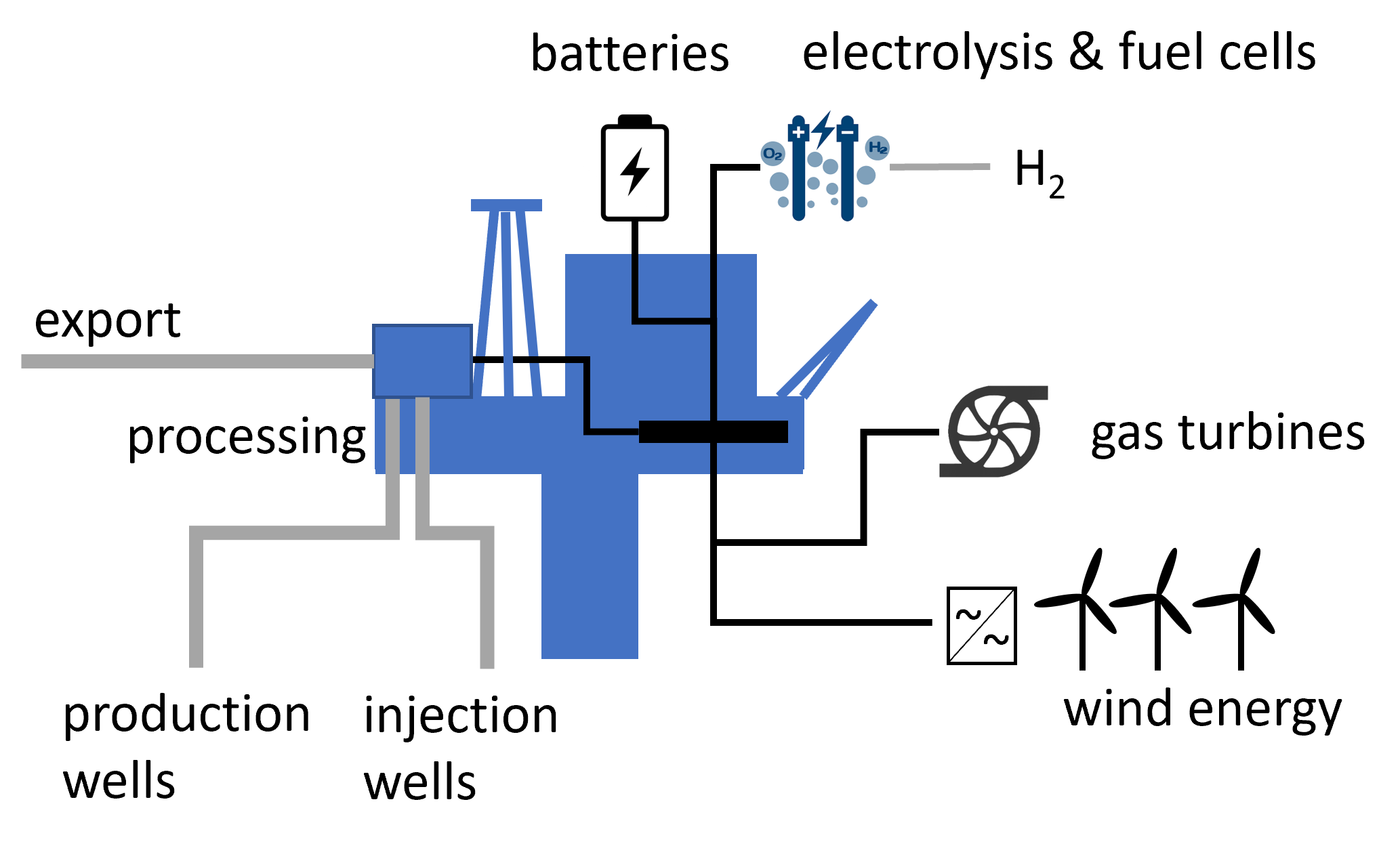}
    \caption{Offshore oil and gas platform with isolated energy system supplied partly by wind energy}
    \label{fig:simplefigure}
\end{figure}

In this paper we analyse an isolated offshore oil and gas platform with wind energy supply in combination with gas turbines and energy storage, as illustrated in Figure~\ref{fig:simplefigure}. The paper has two main parts.
The first part describes an operational optimisation model that has been created as an open-source tool to explore and identify the optimal operational strategy and to analyse integration challenges and benefits with low-emission technologies.
The second part applies this tool to analyse an oil and gas platform. The results give a quantification of \COO emission reductions and other performance indicators with different configurations of wind turbines and energy storage.


\section{An integrated model for energy system operation}

For assessments of how different system designs and operating strategies influence offshore oil and gas platform energy system performance in terms of carbon emissions, fuel costs, oil and gas export rate etc, a new integrated energy system model has been developed and applied in simulation studies.
This energy system model considers multiple energy and fluid flow carriers in  offshore energy systems that may consist of multiple platforms connected together with pipelines and electric power cables. The integrated modelling is essential to capture the links between heat and power demand and oil and gas production and processing. These are important when studying how for example allowing load flexibility in combination with variable wind power supply can reduce emissions whilst maintaining production rates within specified limits.
Details about this model are presented in the following.

\subsection{Model overview}

The  purpose of the model is to allow fast and relatively simple analysis of the energy system, and to provide a convenient way to explore ``what if" type questions. At its core, the model is an operational optimisation model that aims to identify optimal use of available flexible resources.
Some examples of questions the model can address are: How are emissions affected if we add a wind turbine? How does it affect the number of gas turbine startups and shutdowns? What if we allow some flexibility in water injection pumps, how does that influence the need for energy storage?
The emphasis is on the energy supply and distribution system. As mentioned above, energy consumption and its dependence on oil and gas transport and processing is also included, but typically  represented on an aggregated level.

The model is a multi-energy optimisation model, partly inspired by the concept of energy hubs \cite{Geidl4077107,ZHANG20173597}. The approach is familiar from previous studies of integrated energy systems, see e.g. \cite{BAKKEN20071676,Beuzekom7232360}. In most cases, such models are used for investment planning, whereas the application here is operational planning
and comparisons of different system configurations and power management strategies in the operational phase.

\subsection{Devices and networks}
\label{sec:devices-and-networks}

\begin{figure}
\newcommand{\figscale}{0.62}
    \centering
    \begin{tabular}{ccc}
    
    \includegraphics[scale=\figscale]{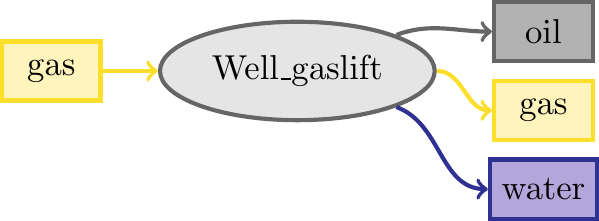} &
    \includegraphics[scale=\figscale]{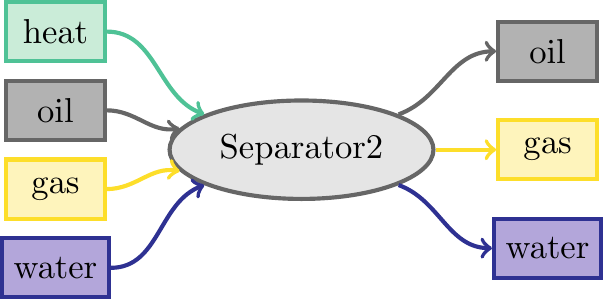} &
    \includegraphics[scale=\figscale]{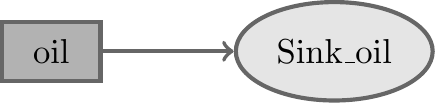} \\

    \includegraphics[scale=\figscale]{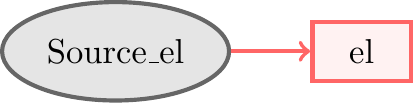} &
    \includegraphics[scale=\figscale]{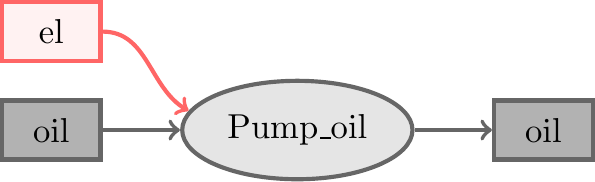} &
    \includegraphics[scale=\figscale]{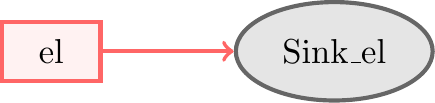} \\
    
    \includegraphics[scale=\figscale]{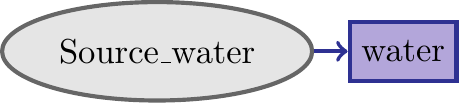} &
    \includegraphics[scale=\figscale]{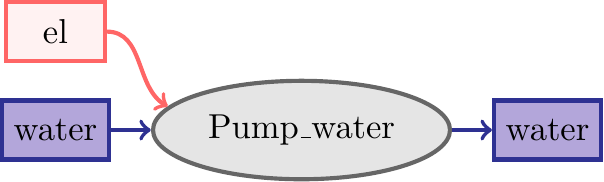} &
    \includegraphics[scale=\figscale]{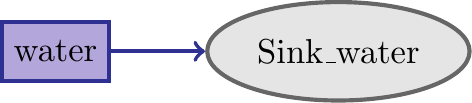} \\

    \includegraphics[scale=\figscale]{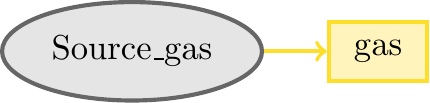} &
    \includegraphics[scale=\figscale]{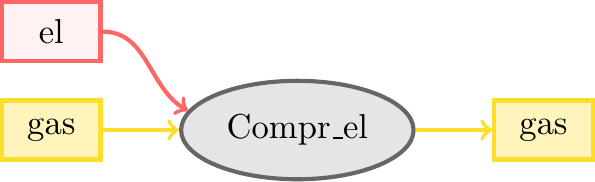} &
    \includegraphics[scale=\figscale]{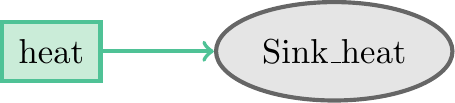} \\
    
    \includegraphics[scale=\figscale]{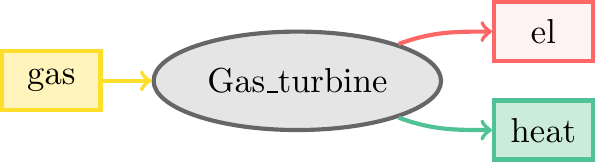} &
    \includegraphics[scale=\figscale]{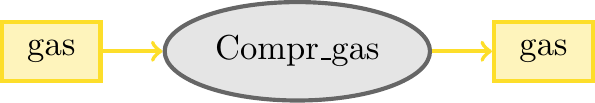} &
    \includegraphics[scale=\figscale]{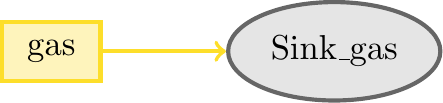} \\

    \includegraphics[scale=\figscale]{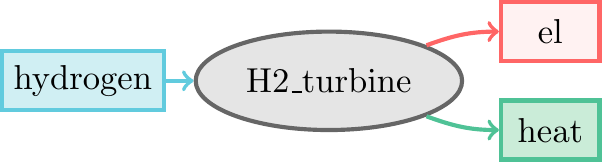} &
    \includegraphics[scale=\figscale]{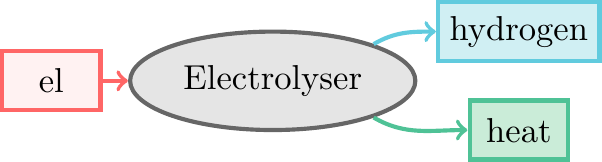} &
    \includegraphics[scale=\figscale]{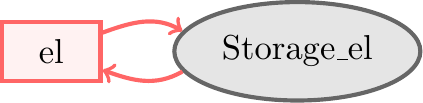} \\
    
    \includegraphics[scale=\figscale]{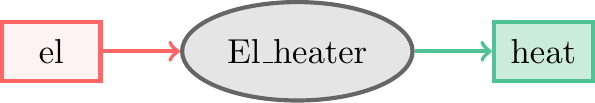} &
    \includegraphics[scale=\figscale]{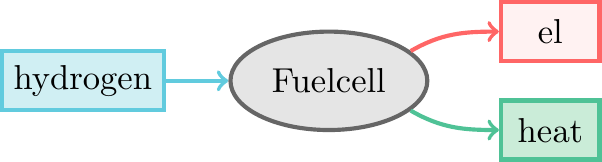} &
    \includegraphics[scale=\figscale]{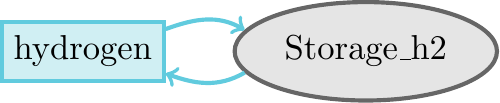} \\
     \\
    \end{tabular}
    \caption{Device types with their in/out flows}
    \label{fig:all_devices}
\end{figure}

The basic building blocks in the model are \emph{devices} and \emph{networks}. 
Devices are all the individual components, such as gas turbine generators, compressors, pumps, auxiliary loads, separator trains, batteries, electrolysers, and others. Generally, devices connect a set of input and output flows of energy/matter. For example, a gas turbine takes gas input and gives electricity and heat output. Mathematical expressions relate these input/output flow quantities:
Different sets of  equations (and inequalities) describe the properties of the individual device types and the relationships between the input and the output.
For computational efficiency, only \emph{linear} relationships are considered in the model.
All device types included in the model, with their respective input/output flows are illustrated in Figure~\ref{fig:all_devices}. The mathematical expressions describing each device type are given in~\ref{sec:deviceconstraints}.

Devices are located at \emph{nodes} that in turn are connected by \emph{edges} (pipelines or cables) into networks.
For each node, there is a set of inlet and outlet \emph{terminals}, one for each network type. And devices and edges are connected via these terminals. Edges are distinguished by which type of flow they carry.

Edges of the same type make up a \emph{network} of that type. That is, there is one network for each carrier, i.e. electricity, heat, oil, gas, water, hydrogen.
The physical energy or matter flow through the networks may be described by different methods.
The simplest is a transport model, where no constraints except upper and lower limits on each edge are included. 
More detailed linear flow models are also possible:
For electricity, the linearised power flow equations relating power flow and nodal voltage could be used\footnote{At present, this is not implemented in the Oogeso tool}, and 
for fluids, the linearised Weymouth  (for gas) or Darcy-Weissbach (for liquids) equations relating flow rates and pressure levels may be used.

Electrical losses may be accounted for via a piece-wise linear loss function relating power flow to power loss. This may have a significant impact in case of long distance electric power connections.

\subsection{Optimisation problem}
\label{sec:optimisationproblem}

The equations and inequalities for all devices and networks are combined with an \emph{objective function} into a mixed-integer linear programming (MILP) optimisation problem. 
Various elements may be included in the objective function, such as e.g. fuel consumption, carbon emission, and other operating costs. 
In general, the objective is to minimise a \emph{penalty}, whose exact interpretation is determined by the user input.

The device and network equations or inequalities mentioned in Section~\ref{sec:devices-and-networks} are \emph{constraints} in the optimisation.
The optimisation problem is solved sequentially, with energy storage, load flexibility and start-up delay linking different time-steps. The problem is solved by looping over time-steps, and updated between each step. 
Time-series profiles describe the variability where relevant, e.g. power availability for wind turbines, wellstream flow, power demand variations etc.

Schematically, the optimisation problem is defined as follows:
\begin{equation}
\label{eq:objectivefunction}
    \begin{split}
        \min \sum_{d\in \mathbb{D}} \Bigl[ &
        \sum_{t\in\mathbb{T}}
        \Bigl\{
        P^d(f_{d,c}^i(t),y^d_\text{on},y^d_\text{prep}) 
        + c^d_\text{start} y^d_\text{start}(t) 
        + c^d_\text{stop}  y^d_\text{stop}(t)
        \Bigr\}
        \\ & 
        + c^d_\text{storage} \delta_E^d
        \Bigr]
        \\
        \text{such that:}
        \\
        & \text{all constraints are satisfied}
    \end{split}
\end{equation}
where
$\mathbb{D}$ is the set of devices,
$\mathbb{T}$ is the set of time-steps in the optimisation horizon (see below),
$f_{d,c}^i$ is the flow of carrier $c$ in or out ($i\in\{\text{in,out}\}$)  of device $d$,
$y^d_\text{on},y^d_\text{prep},y^d_\text{start},y^d_\text{stop}$ are binary variables indicating device on/off status (see below),
$P^d(\dots)$ is a piece-wise linear \emph{penalty function} for device $d$,
$c^d_\text{start}$ is start-up penalty,
$c^d_\text{stop}$ is shut-down penalty,
$c^d_\text{storage}$ is storage depletion penalty,
$\delta_E^d$ is the energy storage filling compared to a target value at the end of the optimisation horizon. The target value could be some value based on long-term storage utilisation planning.
The penalty function $P^d$ may include fuel costs, \COO taxes or other costs (or negative incomes) as a function of input/output flows.
A penalty for storage depletion is relevant when the model includes energy storage devices whose capacity is large compared to the optimisation horizon. Without such a penalty, the optimisation will not see a benefit of charging or filling the storage.

The constraints are spelled out in detail in \ref{sec:allconstraints}. In brief, they include relationships describing the devices, energy or matter balance in each terminal, network flow equations, and reserve power requirements.

\subsection{Start-stop logic and rolling horizon}

\begin{figure}
    \centering
    \includegraphics[scale=0.5]{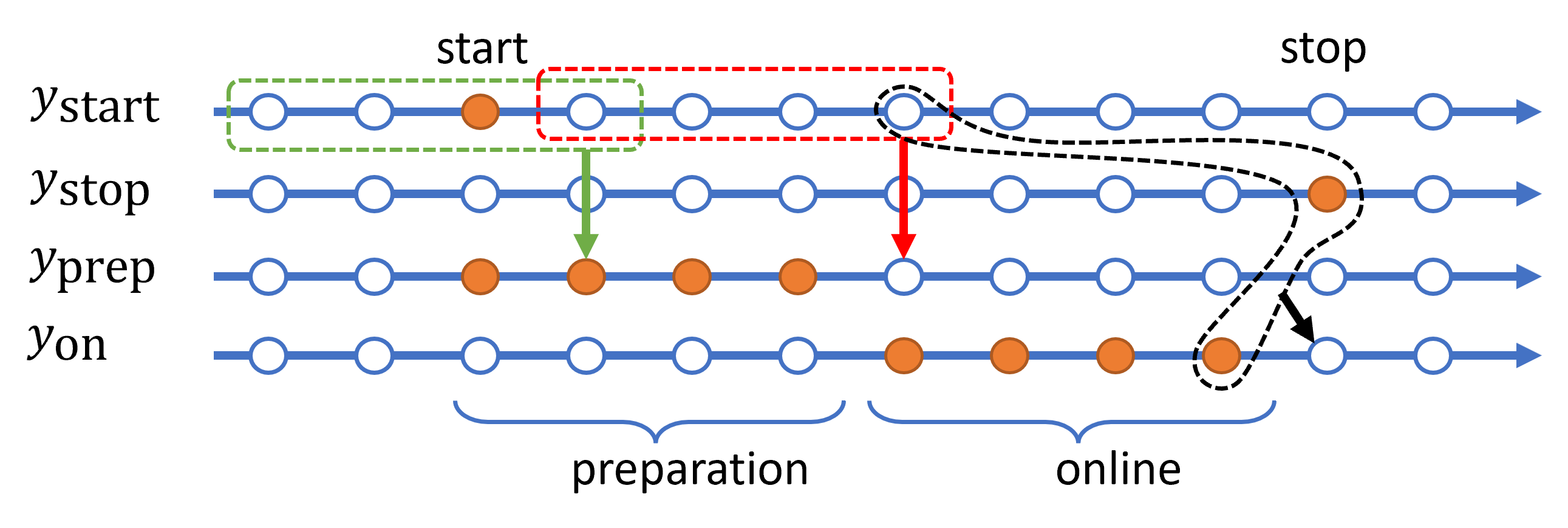}
    \caption{Start--stop logic. Illustrative example with a preparation time of 4 time-steps. Colouring of the dots indicate binary value 1 (orange) or 0 (white). The control variables $y_\text{start}$ and $y_\text{stop}$ determine the preparation and on/off status.}
    \label{fig:startstop}
\end{figure}

Optimised start and stop of gas turbine generators, and possibly other devices, can have a significant impact on the energy system performance. 
Since there is some preparation time from start-up decision to actual power output and potential limitations on ramping rates, forecasts and planning ahead is important.
To keep track of on/off and activation status, four binary variables are used  
\cite{doi:https://doi.org/10.1002/9780471225294.ch2}.
The decision variables are $y_\text{start}$ and $y_\text{stop}$, which are 1 if a decision to start or stop is made. 
The two derived variables $y_\text{prep}$ and $y_\text{on}$  indicate whether the device is in startup preparation or online respectively. 
This is illustrated in Figure~\ref{fig:startstop} for an example with preparation time of 4 time-steps: The device is in preparation if it was started in the last 4 time-steps, and online if it was started 4 steps previously or online in the previous step and not stopped in the present step (see equation~\ref{eq:sartstop} and \ref{eq:startdelay}).
A gas turbine generator consumes fuel also in the preparation phase, but provides electric power only when it is in the online state.

\begin{figure}
    \centering
    \includegraphics[scale=0.5]{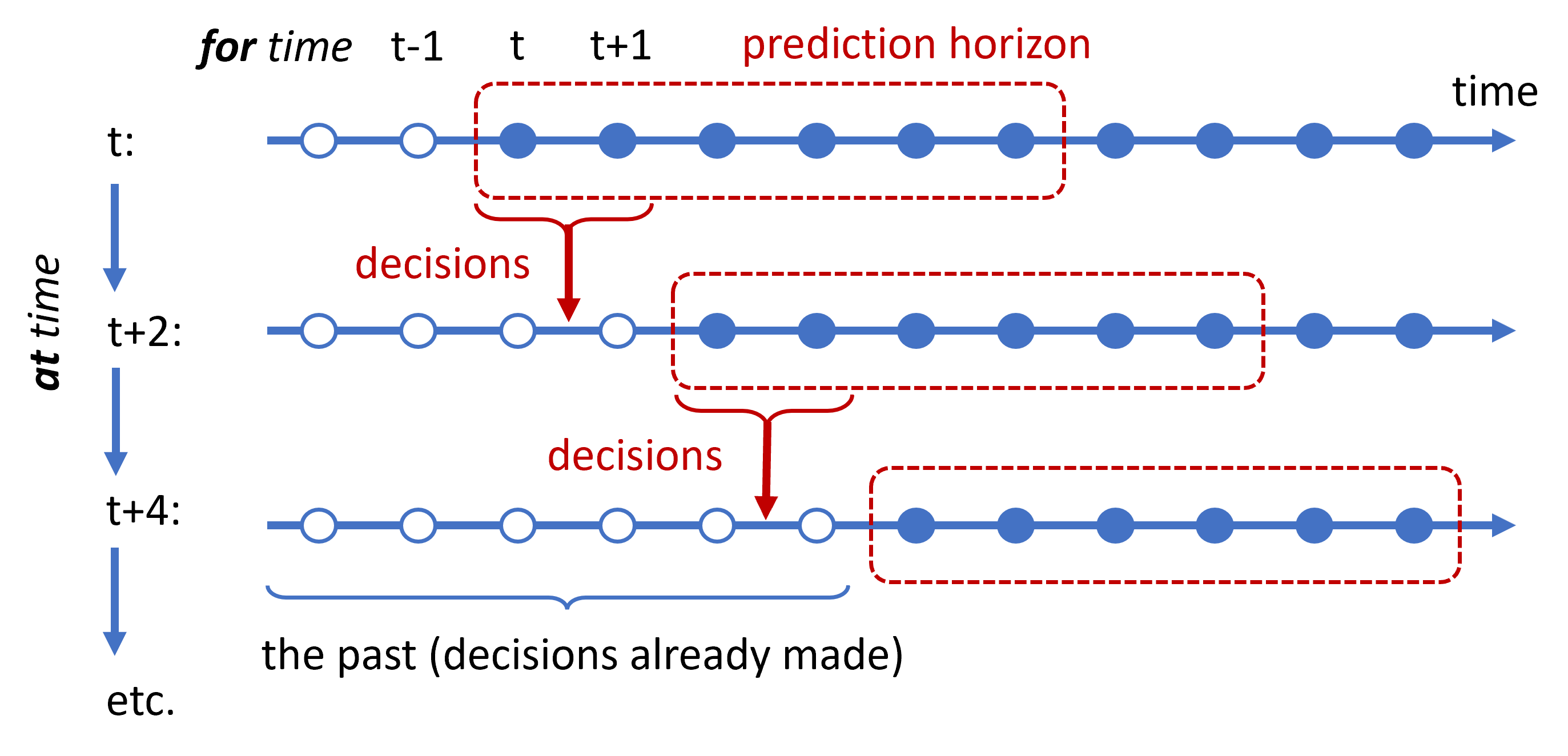}
    \caption{Rolling horizon optimisation with a prediction horizon of 6 time-steps, repeated every 2 time-steps (example)}
    \label{fig:rollinghorizon}
\end{figure}

A rolling horizon optimisation is used to account for start-up delays, short-term flexibility in energy storage and energy demand, and the availability of improved forecasts closer to real-time. This means that the optimisation is repeated at regular intervals as time moves forward. This is illustrated in Figure~\ref{fig:rollinghorizon}. 
\emph{At} a given time-step $t$, the system is optimised \emph{for} a certain prediction horizon $(t,\dots, t+t_H)$, where $t_H$ is the number of time-steps in the prediction horizon.  
Variable values for the first time step(s) represent decisions while values for later time steps are preliminary plans to be fixed or adjusted in subsequent iterations.
In this figure, the optimisation is repeated every 2 time-steps, and the prediction horizon is 6 time-steps.

\subsection{The Oogeso tool}

The above generic modelling concepts have been implemented as a new open source Python package named \emph{Offshore Oil and Gas Energy System Optimisation} tool, or \emph{Oogeso} for short \cite{oogeso}.

This tool optimises system operation in the presence of variability and different sources of flexibility. Relevant decision variables are for example when to start-up or shut-down a gas turbine, when to charge or discharge a battery, or when to invoke load flexibility.
In principle, it is an operational planning tool that identifies optimal set-points that can be fed into  an energy management system. 
The main intended use, however, is to simulate offshore multi-energy  systems and compute fuel consumption,  greenhouse gas emissions, number of gas turbine starts and stops, and other key performance indicators, and use these to assess and compare simulation cases with different system designs or parameter choices reflecting different operating strategies. 
In this context, the purpose of the optimisation is to obtain a realistic operation of the system as part of a system analysis, rather than to identify optimal set-points in themselves.

It should be noted that the Oogeso tool is an \emph{energy system} analysis tool, with only a simple representation of the oil and gas processing system, and that it does not include a reservoir model.
However, it includes the connection between flow rates and energy demand for pumps and compressors which are crucial to address the potential for energy demand flexibility.
The Oogeso tool allows devices and networks to be arranged in any way, making it suitable to analyse a broad range of different configurations without too much effort on the modelling side.

\subsection{Online power reserve}

An important system parameter affecting energy system operation is the required online power reserve (spinning reserve). This is power capacity reserved to balance unforeseen variations in demand and supply due to for example a motor start-up or wind forecast error. In traditional offshore power systems, only gas turbine generators provide reserve. In general, however, unused generator capacity, potential storage discharge capacity and potential load reduction may all contribute to the reserve. 
The reserve in an offshore energy system is typically not sufficient to cover all demand in case of a generator failure or other similar rare events. In those cases, frequency-based load shedding will be activated. 
The Oogeso model considers operational planning close to real-time, but not such real-time control actions.

\section{Simulation setup}

This section presents results from simulations of a relevant offshore oil and gas platform using the   Oogeso tool described in the previous chapter.

The aim of the study is to assess and compare different alternatives for carbon emission reductions by adding wind energy supply to an existing platform, with and without energy storage. Additionally, this study demonstrates the applicability of the Oogeso analysis tool on a relevant case.

\subsection{The LEOGO reference platform}

The Low Emission Oil and Gas Open (LEOGO) reference platform \cite{leogodata} has been used as a basis for demonstrating the model and to analyse some alternatives for reducing carbon emissions.
The LEOGO platform is a hypothetical but realistic case that is representative of a Norwegian offshore oil and gas production platform. It is well suited for our present purpose since the specification and associated data is public.

In brief, the LEOGO platform represents an oil and gas platform with production rates of
\SI{8 600}{Sm^3/day}\footnote{\si{Sm^3} denotes  cubic metres under standard conditions.} oil,
\SI{4.3}{mill.~Sm3/day} gas, and 
\SI{13 000}{Sm3/day} water. The gas-to-oil ratio is 500 and the water cut is 0.6.
Wellstream transport  is supported by gas lift and reservoir pressure is maintained by water injection. Sea water is used in addition to production water for the water injection.
Extracted oil and gas is exported via pipelines to a nearby platform.
The main energy consumers are the gas compressors, oil export pumps, and water injection pumps. 
Electricity demand is about \SI{40}{MW}, in the base case supplied by three gas turbines with active power rating of \SI{21.8}{MW}. 
Heat demand is about \SI{8}{MW}, supplied by gas turbine heat recovery.

\begin{figure}
    \centering
    \includegraphics[width=\columnwidth]{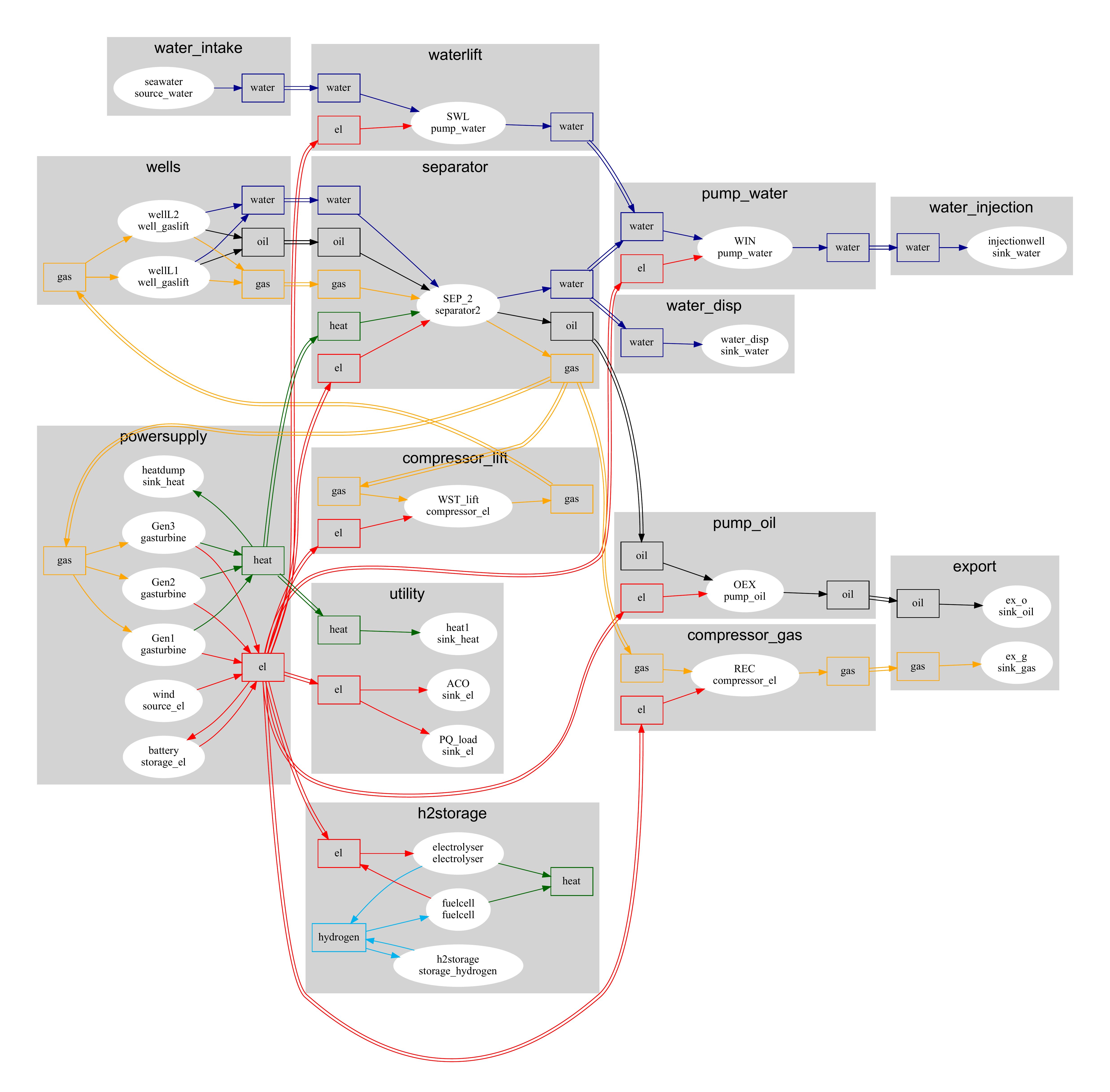}
    \caption{Overview of the LEOGO model implemented in Oogeso. Edge colours represent different energy carriers. Gray boxes are nodes and white ellipses are devices.}
    \label{fig:leogo}
\end{figure}

The platform as represented in the Oogeso tool is illustrated in Figure~\ref{fig:leogo}.

\subsection{Energy demand and supply forecasts}

In an energy management system with real-time operational optimisation, the best available forecasts for energy demand and supply would be used in each optimisation run.
In a \emph{simulation} setup, this is a bit different as the forecasts are given by input time-series up-front. To account for forecast errors when looking farther ahead, we use two different time-series for wind power: A \emph{forecast} which represents the best prediction some time ahead, and a \emph{nowcast} which represents the improved prediction for the next few minutes.
In our case, the forecast has been generated from the nowcast simply by the addition of Gaussian noise.

In the simulations presented below, the nowcast was used for the first 10 minutes and the forecast was used for the rest of the optimisation horizon.
This is a simplification of the real-world situation, but enables us to see the effect of  forecast uncertainty to some extent.

\begin{figure}
    \centering
    \includegraphics[scale=0.5]{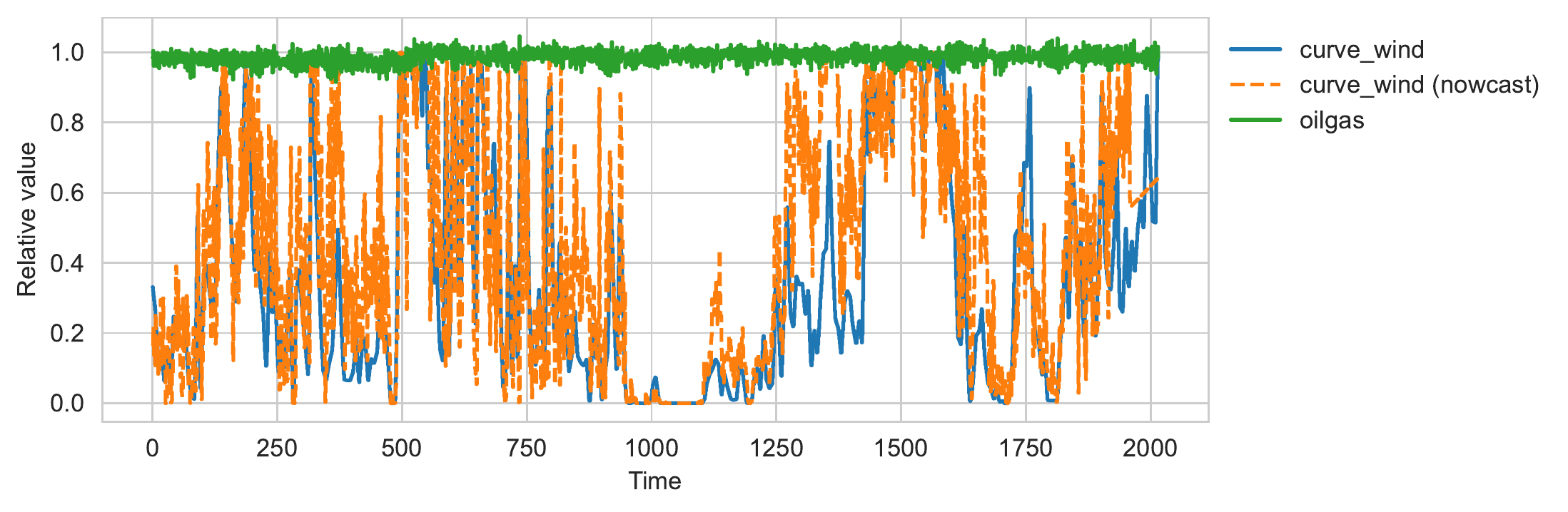}
    \caption{Time series profiles for wind forecasts and oil and gas production (wellstream flow rate)}
    \label{fig:profiles}
\end{figure}

Simulations were run for a one-week period, using wind speed data from the first week of April 2020, see Figure~\ref{fig:profiles}.

\subsection{Simulation cases}

Four simulation cases, summarised in Table~\ref{tab:cases}, have been defined. The base case corresponds to a typical present-day situation with power supply from local gas turbines. The three variations expand on this case with the addition of wind power, battery storage and hydrogen storage (via electrolyser and fuel cells). The battery storage is fairly small, but provides sufficient reserve power that a gas turbine may be allowed to switch off. The hydrogen storage is large enough to supply the full power demand for about 4 days. 
These simulation cases are chosen to illustrate some generic performance differences, but not selected in detail.

\begin{table}
    \centering
    \caption{Simulation cases}
    \label{tab:cases}
    \begin{tabular}{p{0.2\textwidth}p{0.2\textwidth}p{0.2\textwidth}p{0.2\textwidth}}
    \hline
    Base case & Variation A & Variation B & Variation C\\
    \hline
    3  gas turbines
    & {\raggedright
    3 gas turbines\\
    24~MW wind power\\}
    &
    {\raggedright 3 gas turbines\\
    24~MW wind power\\
    4~MW/4~MWh battery}
    & {\raggedright
    3 gas turbines\\
    96~MW wind power\\
    Hydrogen storage}
    \\
    \hline
    \end{tabular}
\end{table}

\begin{table}
\centering
\caption{Main simulation parameters}
\label{tab:simulationparameters}
\begin{tabular}{lr}
\hline
Parameter                              & Value   \\ \hline
Time resolution                        & 5 min   \\
Time between each optimisation         & 30 min  \\
Planning horizon                       & 120 min \\
Reserve required                       & 5 MW    \\
Gas turbine generator max power        & 21.8 MW \\
Gas turbine generator min power        & 3.5 MW  \\
Gas turbine generator startup time     & 30 min  \\
\hline
\end{tabular}
\end{table}

For the simulations, time-steps of 5 minutes were used. This resolution is high enough to  account for wind power variability and gas turbine generator start-up delays.
Further key input parameters are given in Table~\ref{tab:simulationparameters}. The full data and simulation settings are provided with the LEOGO dataset \cite{leogodata}.

\section{Results}

\begin{figure}
    \centering
    \includegraphics[scale=0.56]{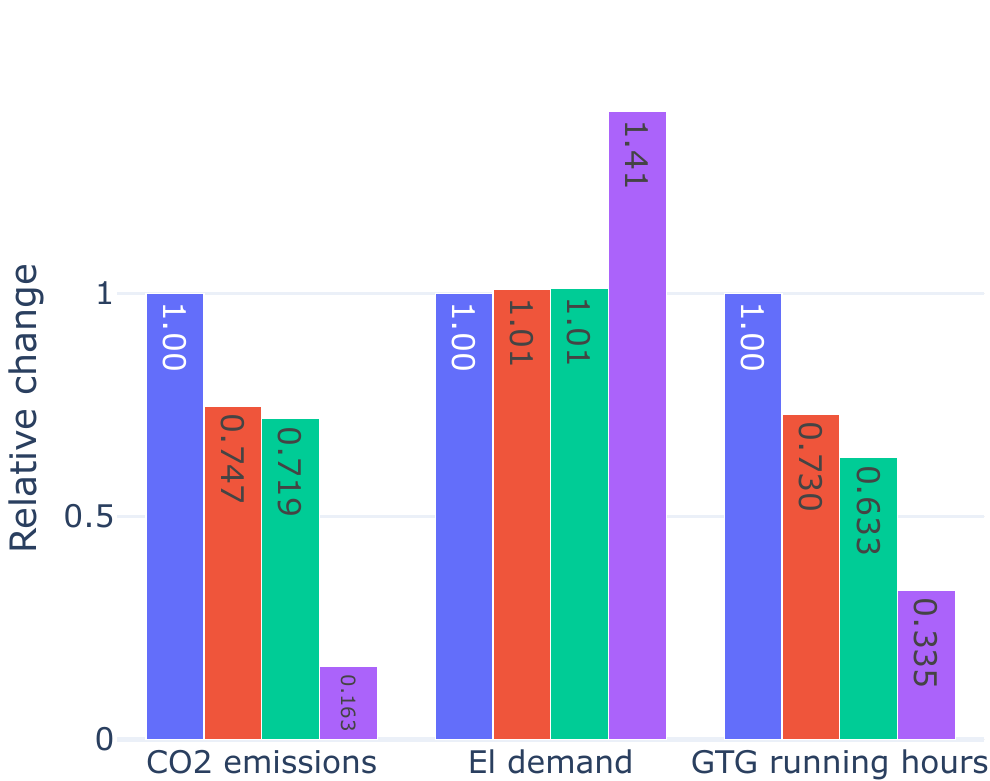}
    \includegraphics[scale=0.56]{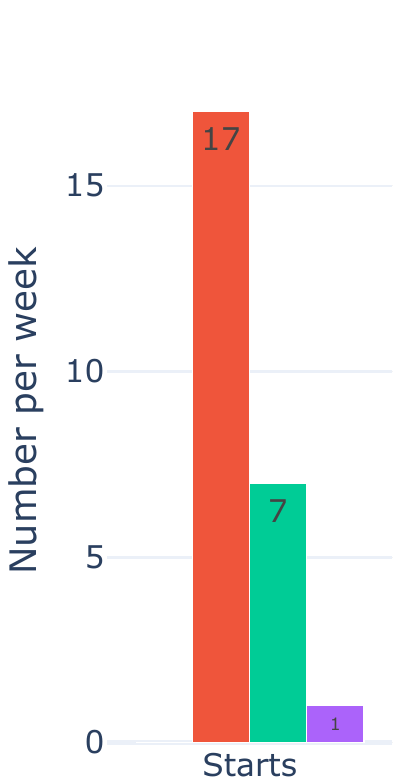}
    \includegraphics[scale=0.56]{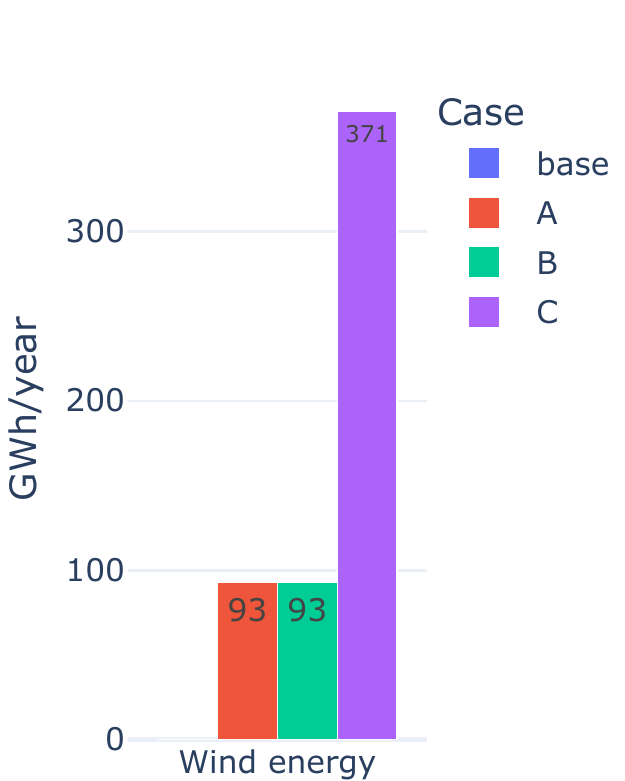}
    \caption{Simulation results}
    \label{fig:kpis}
\end{figure}

An overview and comparison of simulation results is provided in Figure~\ref{fig:kpis}.
The main observations from the simulations are elaborated in the following.

\subsection{Emissions}

\begin{figure}
    \centering
    \includegraphics[scale=0.56]{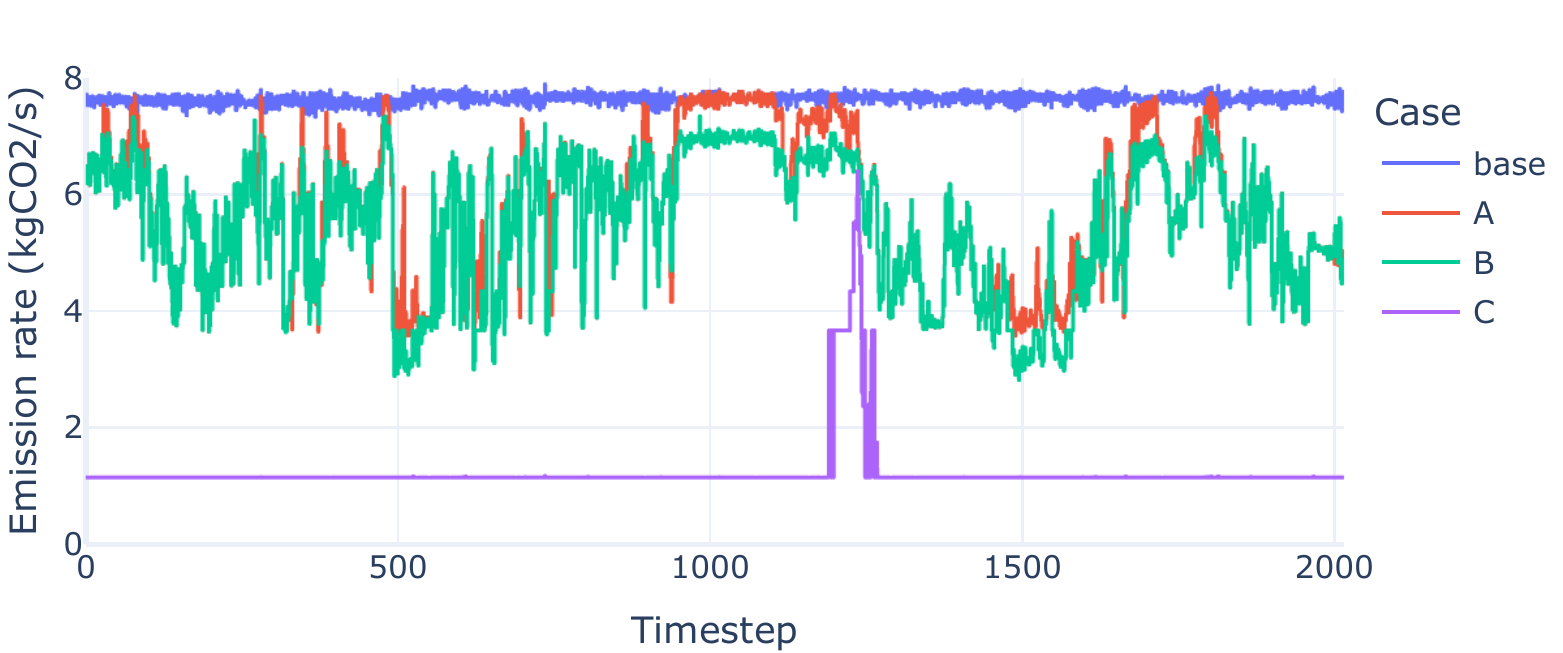}
    \caption{\COO emission rates}
    \label{fig:emissionrate}
\end{figure}

\COO emission rates for the different simulation cases are compared to the base case in the first plot in Figure~\ref{fig:kpis}, and absolute numbers and time-variation are shown in Figure~\ref{fig:emissionrate}.
In the base case, a more or less constant power demand is supplied by three gas turbine generators, and the emission rate is steady at close to \SI{8}{kgCO_2/s}. Wind power reduces gas turbine fuel usage and therefore emissions. The reduction is \SI{25}{\%} in variation A and \SI{28}{\%} in variation B. The slightly better performance in variation B is because the reserve provided by the battery makes it possible to switch off an extra gas turbine, thus allowing gas power generation at higher efficiency.

\begin{figure}
    \centering
    \includegraphics[scale=0.5]{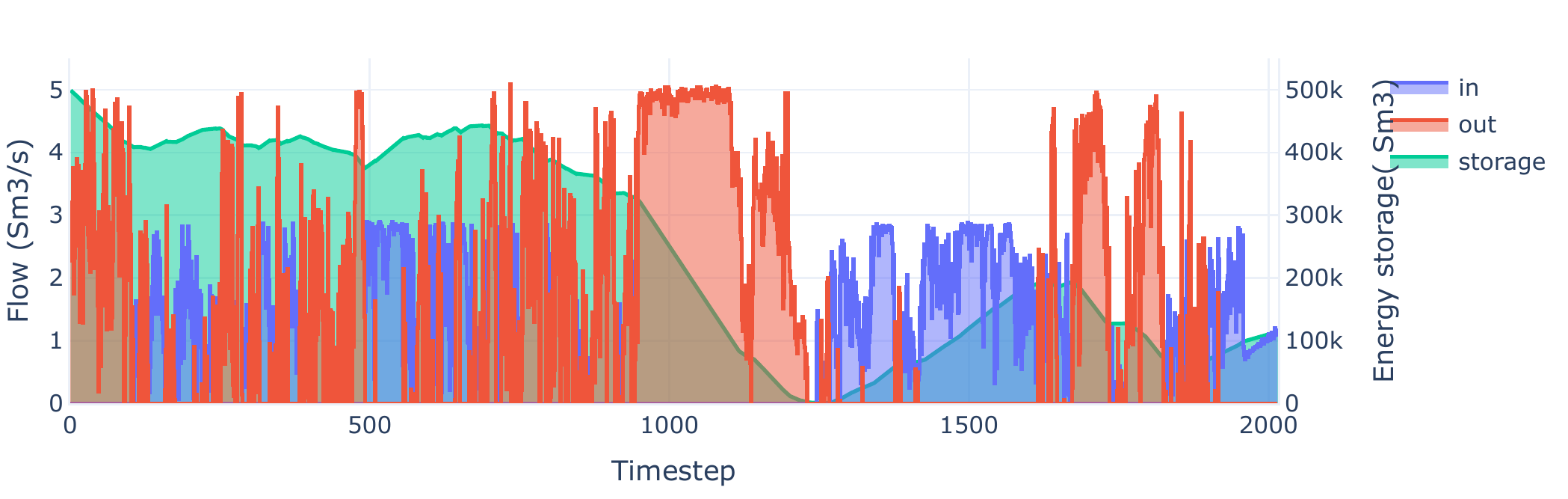}
    \caption{Hydrogen storage utilisation (Variation C). Flow in and out (left hand axis) and storage filling level (right-hand axis)}
    \label{fig:h2storage}
\end{figure}

In Variation C, the wind power capacity has been increased to 96~MW with a total wind energy production similar to the energy demand. 
The electricity demand is increased by \SI{41}{\%} because of the electrolyser which produces hydrogen when there is excess wind power.
Since there is no other heat source, one gas turbine is still needed to supply the heat demand and giving rise to \COO emissions. 
If waste heat recovery from the electrolyser and fuel cell or an electric heater were included instead, this would not be needed and the emission rate could be reduced to almost zero. The jump in emissions at time-step around 1200 is because of sustained low wind conditions where the hydrogen storage has run empty, see Figure~\ref{fig:h2storage}.
Therefore more power and even a second gas turbine is required for a short time. 


%

\subsection{Gas turbine utilisation}

Gas turbine running hours are compared in the third plot in Figure~\ref{fig:kpis}, and usage goes down as wind energy share increases. In this case, it is interesting to note that variation B with a battery has a significant improvement compared to variation A. The same is seen when comparing the number of starts (and stops) in the fourth plot. 
The reason for this is the \SI{4}{MW} online power reserve provided by the battery, which lowers the needed reserve from the gas turbines from \SI{5}{MW} to \SI{1}{MW} (sum). The gas turbines can therefore be run closer to maximum power, allowing one or two gas turbines to be shut-down more often in case B than in A.
Both the gas turbine running hours and the number of starts and stops is therefore considerably less in variation B than in variation A.
As noted above, in variation C, a single gas turbine is needed for the heat demand with minimal need for starts and stops.

It is clear from this that the relationship between the power demand, the  gas turbine power capacity and the reserve requirement has a significant impact on emission reduction and number of gas turbine starts and stops. To obtain the maximum benefits, selecting the right capacities in the design phase and tuning of operational strategies is therefore important.

\begin{figure}
    \centering
    \includegraphics[scale=0.2]{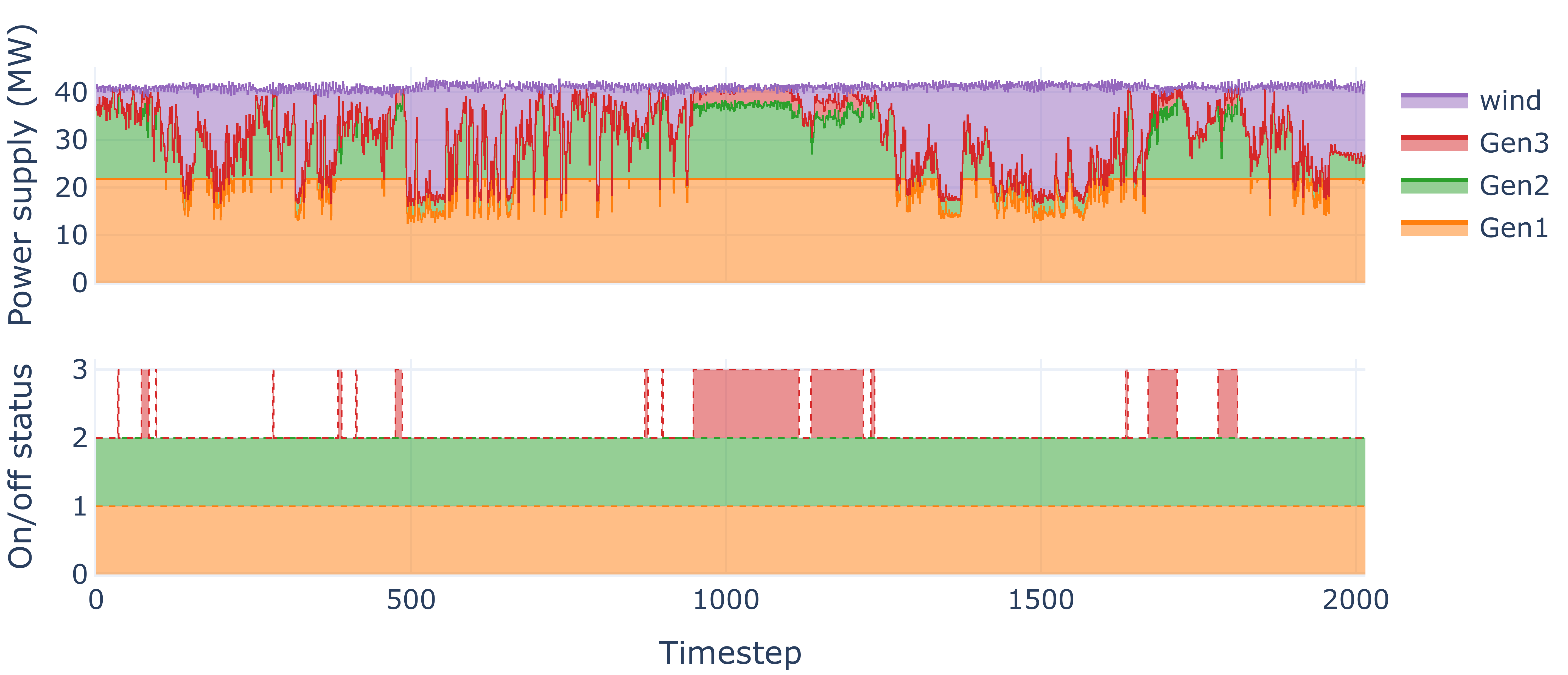}
    \caption{Power supply and generator start/stop status (Variation A)}
    \label{fig:powersupply}
\end{figure}

Figure~\ref{fig:powersupply} shows a time-series of power supply and start/stop status of the gas turbine generators in Variation A simulations. The gas turbines balance the variable wind power to cover the electricity demand which is just over 40 MW. The third gas turbine generator is only needed when wind power is low.
Note that the power sharing between the online generators is not realistically represented in this plot, and is an artefact of the linear modelling: In reality, generator controls would ensure equal loading of the online generators.

\subsection{Reserve}

\begin{figure}
\centering
\begin{subfigure}{.45\columnwidth}
  \centering
  \includegraphics[width=0.95\linewidth]{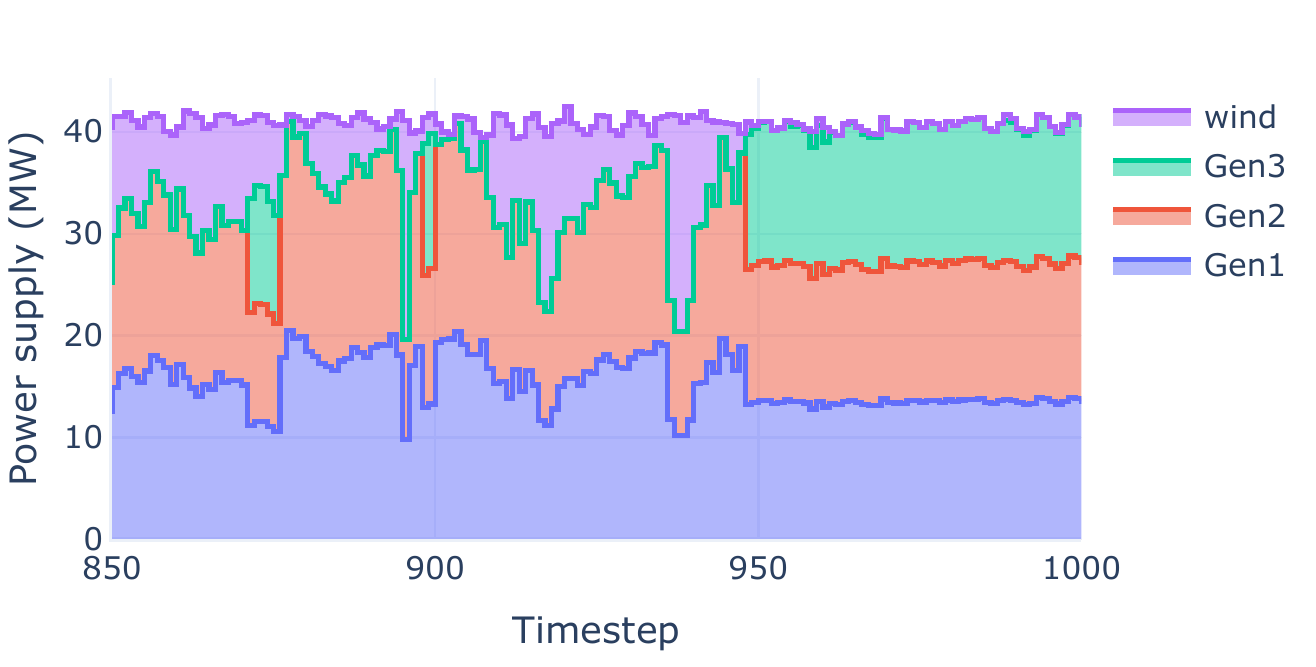}  
  \caption{Variation A -- power supply}
\end{subfigure}
\begin{subfigure}{.45\textwidth}
  \centering
  \includegraphics[width=.95\linewidth]{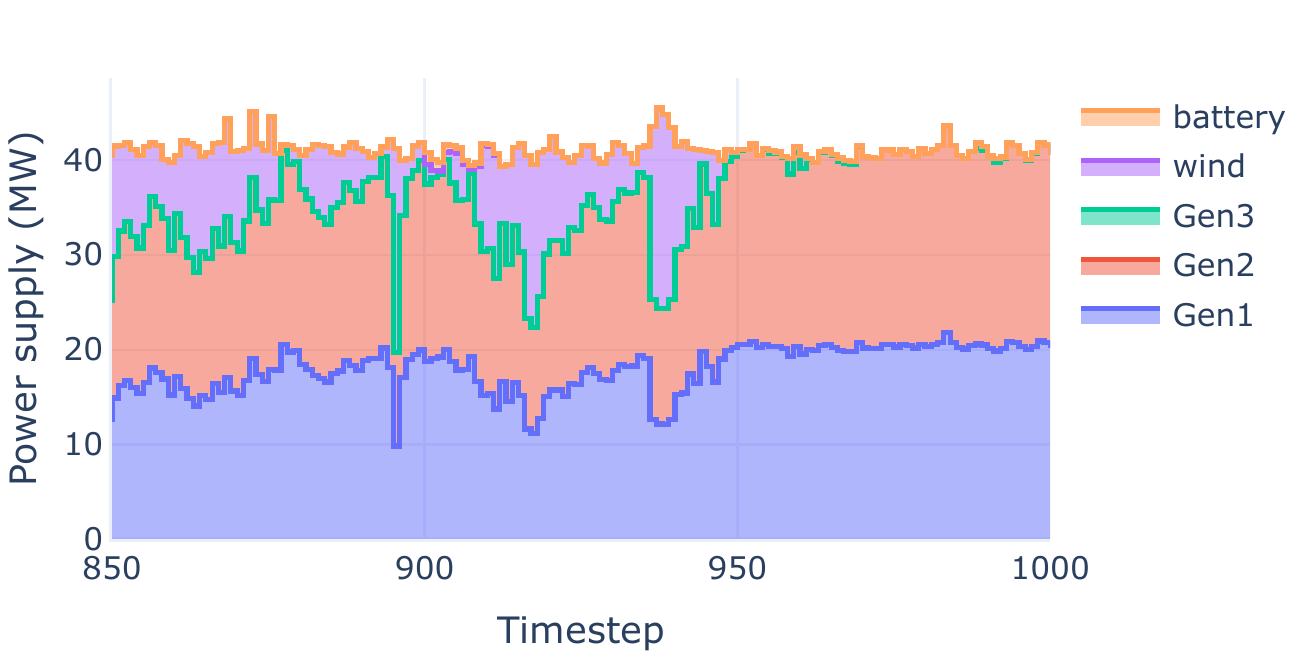}  
  \caption{Variation B -- power supply}
\end{subfigure}
\begin{subfigure}{.45\columnwidth}
  \centering
  \includegraphics[width=0.95\linewidth]{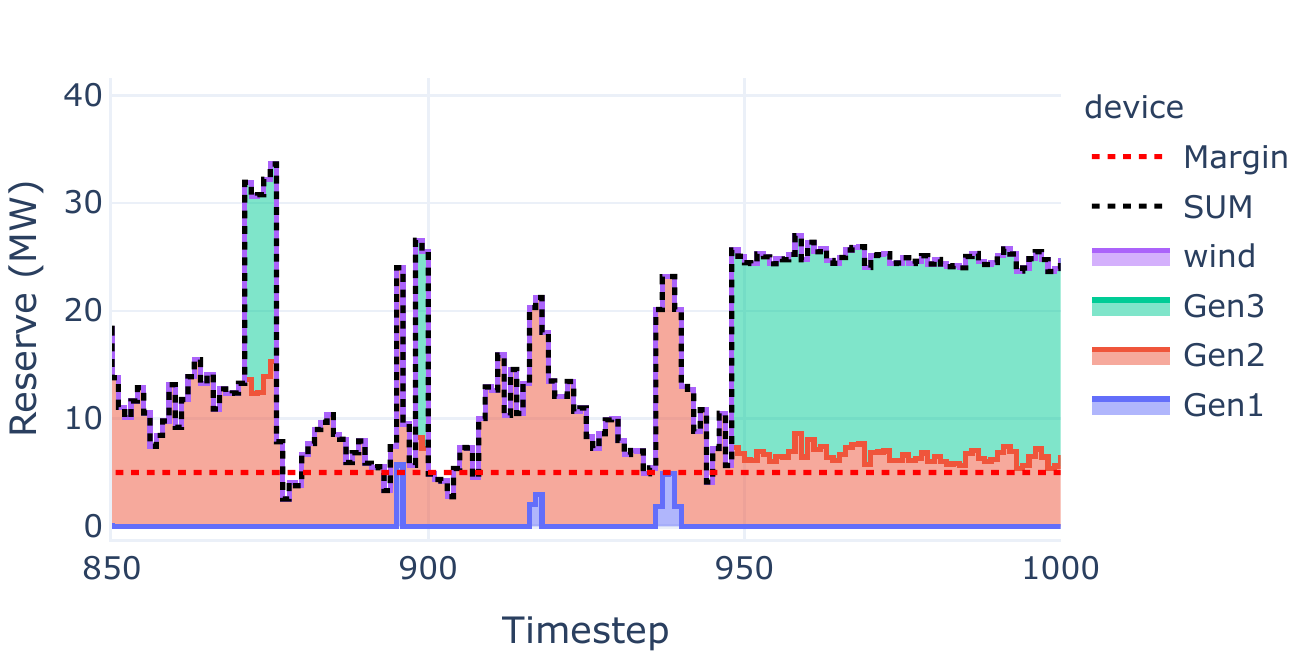}  
  \caption{Variation A -- reserve}
\end{subfigure}
\begin{subfigure}{.45\columnwidth}
  \centering
  \includegraphics[width=0.95\linewidth]{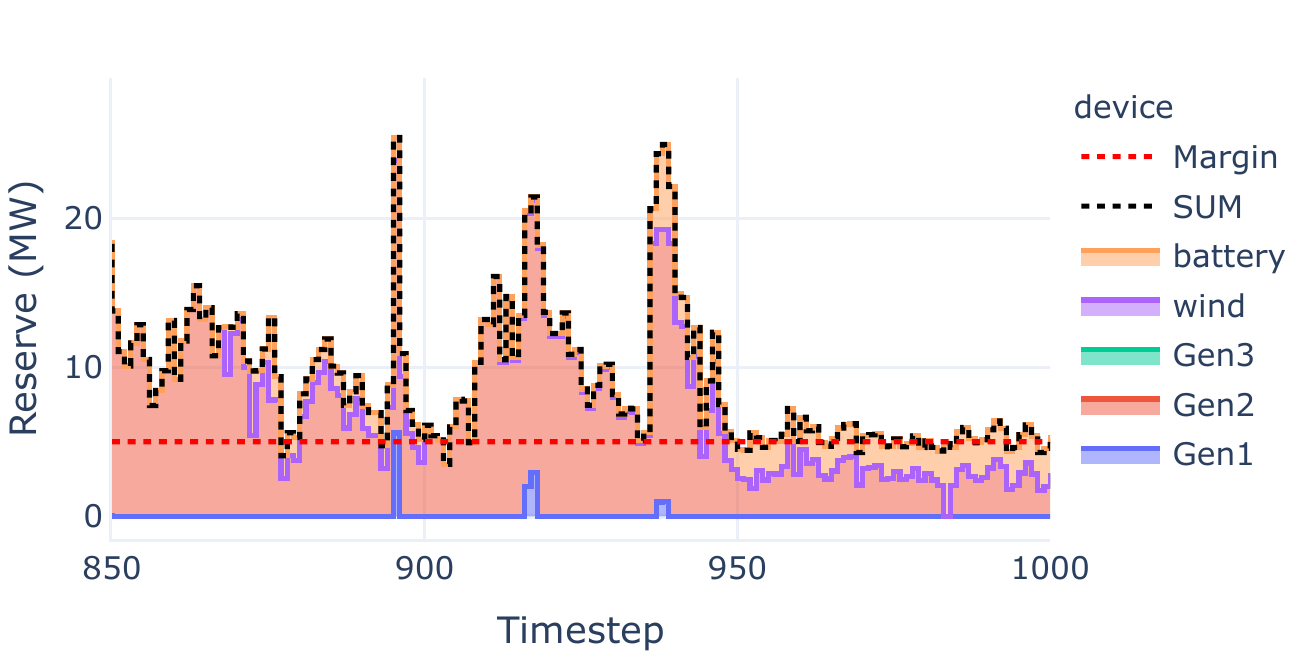}  
  \caption{Variation B -- reserve}
\end{subfigure}
\caption{Comparison of power output (top) and available resrve (bottom) in Variations A and B. The battery in Variation B contributes to reserve and avoids the need for starting up an exta gas turbine at timestep 950.}
\label{fig:reserve_illustration}
\end{figure}

The impact of the battery on the reserve requirement is illustrated in more detail in Figure~\ref{fig:reserve_illustration} which shows power supply and available reserve from each of the power sources in Variation A and Variation B.
The importance of the reserve requirement is clear. The dotted red line shows the required margin of 5~MW, and the total online power reserve must be above this limit. Around time-step 950, the wind drops to zero, but even though the two gas generators has a combined capacity of 43.6 MW, a third power source is needed to provide sufficient power reserve. In Variation A this is done by starting up a third generator, whereas in Variation B, stored energy and power capacity of the battery provides the required reserve.
In fact, in these simulations with fairly constant demand around 41 MW, we can easily deduce that the battery or a third generator is needed when the wind power forecast is less than about 2.4~MW.

\section{Conclusion}
\label{sec:conclusion}

The Oogeso open-source optimisation and simulation tool for analysis of energy system operation has been presented and demonstrated for a relevant offshore oil and gas platform case. 
It has proven useful to assess the effects of modifications in system configurations and operating parameters, such as e.g. wind power capacity, reserve requirement, energy storage capacity, and  their  impact  on  \COO emissions  and  other  performance indicators.

The results have quantified the benefits of adding wind turbines and energy storage to reduce \COO emissions whilst satisfying electricity and heat demand within specified reserve safety margins. 
The case variations have been selected to illustrate key capabilities of the modelling approach and do not represent optimised alternatives. The details of the simulation results should be considered in light of this.

The work presented in this paper opens multiple directions for follow-on work. 
On the modelling side, planned future work will consider demand-side flexibility and how this can be efficiently represented in the operational optimisation model. Utilising load flexibility will be important to minimise the need for costly energy storage alternatives. Load flexibility may also lower the power reserve requirement. 
Another improvement would be to use forecast uncertainty and an option to determine the reserve requirement dynamically: If the forecast is uncertain, more reserve is needed.
Although the Oogeso tool has  been made for analyses of offshore oil and gas installations, the model is generic and can be used also for other types of multi-energy systems. To better represent thermal systems, the fluid temperature should be explicitly included, as well as thermal losses.

\section*{Acknowledgements}

The work reported here has received financial support by the Research Council of Norway through PETROSENTER LowEmission (project code 296207) and through the OFFLEX project (project code 319158).
Insightful comments on the manuscript by colleague Til K Vrana are greatly appreciated.

\appendix

\section{Oogeso Optimisation problem constraints}
\label{sec:allconstraints}

The following sections describe in detail the equations and inequalities governing the energy system as represented in the Oogeso model, i.e. all the constraints in the optimisation problem that was outlined in Section~\ref{sec:optimisationproblem}.
The constraints are grouped into constraints associated with devices, with networks (nodes or edges) and global constraints.

For questions regarding the exact implementation that are not clear from this presentation, the reader is pointed to the open-source code \cite{oogeso}.

\subsection{Device constraints}
\label{sec:deviceconstraints}

Devices are governed by some relationships that are generic and common for all device types, and some that are specific to each device type.
The generic constraints, corresponding to 
max/min flow (e.g. power output) (\ref{eq:max_min}),
ramp rate limits (\ref{eq:ramprate}), and
on/off status (\ref{eq:sartstop} and \ref{eq:startdelay}),
are:


\begin{align}
	y_\text{on} f^\text{min} pr(t) & \le f \le y_\text{on} f^\text{max} pr(t)
	\label{eq:max_min}
	\\
	-r_{-}^\text{max} &\le \frac{f(t) - f(t-1)}{ f^\text{max}\Delta t} \le r_{+}^\text{max}
	\label{eq:ramprate}
	\\
	y_\text{on}(t) - y_\text{on}(t-1) &= y_\text{start}(t-t_s)-y_\text{stop}(t)
	\label{eq:sartstop}
	\\
	y_\text{prep}(t) &= \sum_{\tau=0}^{t_s-1} y_\text{start}(t-\tau)
	\label{eq:startdelay}
\end{align}
where
$t$ is the time-step,
$f$ is a flow variable whose definition depends on the device type (specified below),
$f^\text{max}$ and $f^\text{min}$ are maximum and minimum flow limits,
$pr(t)$ is a time-series profile  (or 1 if no profile is specified
$y_\text{on}$ and $y_\text{prep}$ are binary variables indicating whether the device is on or in preparatory (startup) phase,
$y_\text{start}$ and $y_\text{stop}$ are bindary variables indicating wheter the device is starting or stopping,
$t_s$ is the number of timestep the device spends in preparatory phase,
$\Delta t$ is the timestep length, 
and
$r_\pm^\text{max}$ is maximum up and downwards ramping limit.

The constraints specific to each type are given in the following.


\paragraph{Well}
The well is the source for the combined oil/gas/water wellstream from the reservoir.  The model allows for gas injection, whose purpose is to assist the wellstream flow. The set of governing equations are:
\begin{align}
    f &=  (f^\text{out}_\text{gas}-f^\text{in}_\text{gas})
        +f^\text{out}_\text{oil} + f^\text{out}_\text{water}
    \\
    p^\text{gas}_\text{out} = p^\text{oil}_\text{out} = p^\text{water}_\text{out} &= \hat{p}_\text{separator}
    \\
    f^\text{out}_\text{water} & = \frac{r_w}{1 + R_s(1 - r_w)} f
    \\
    f^\text{out}_\text{oil} & = \frac{1 - r_w}{1 + R_s(1 - r_w)} f
    \\
    f^\text{out}_\text{gas}-f^\text{in}_\text{gas} & = \frac{R_s(1-r_w)}{1 + R_s(1-r_w)} f
    \\
    f^\text{in}_\text{gas} &= r_\text{inj} f^\text{out}_\text{oil}
    \\
    p^\text{in}_\text{gas} &= \hat{p}_\text{inj}
\end{align}
where 
$f$ is the wellstream flow from the reservoir into the well,
$\hat{p}_\text{separator}$ is the separator inlet pressure and assumed fixed.
$R_s$ is the gas over oil ratio (GOR),
$r_w$ is the water cut (WC) ratio,
$r_\text{inj}$ is the gas injection ratio, which is assumed fixed,
and
$\hat{p}_\text{inj}$ is the gas injection pressure, which is also assumed fixed.

\paragraph{Separator}
The separator train is the device that splits the composite wellstream into its individual gas, water and oil components. The representation here is a simple model where the heat and electricity demand are assumed proportional to the flow.
The following equations describe separator devices:
\begin{align}
    f^\text{out}_\text{oil} &=  f^\text{in}_\text{oil}
    \\
    f^\text{out}_\text{gas} &=  f^\text{in}_\text{gas}
    \\
    f^\text{out}_\text{water} &=  f^\text{in}_\text{water}
    \\
    p^\text{gas}_\text{out} &=\hat{p}^\text{gas}_\text{sep,out}
    \\
    p^\text{oil}_\text{out} &=\hat{p}^\text{oil}_\text{sep,out}
    \\
    p^\text{water}_\text{out} &= \hat{p}^\text{water}_\text{sep,out}
    \\
    f^\text{in}_\text{heat} &= r_\text{heat} (f^\text{in}_\text{oil}+f^\text{in}_\text{gas}+f^\text{in}_\text{water})
    \\
    f^\text{in}_\text{el} &= r_\text{el} (f^\text{in}_\text{oil}+f^\text{in}_\text{gas}+f^\text{in}_\text{water})
\end{align}
where
$r_\text{heat}$ is the heat demand factor,
$r_\text{el}$ is the electricity demand factor,
$\hat{p}^\text{gas}_\text{sep,out}$, $\hat{p}^\text{oil}_\text{sep,out}$, $\hat{p}^\text{water}_\text{sep,out}$ are nominal (assumed fixed) separator outlet pressures,

\paragraph{Compressor}
Compressors are used to increase the pressure of natural gas, needed for gas lift and gas transport (export).
The equation expressing the compressor power consumption is derived from the work
done compressing the gas in an adiabatic process, assuming the ideal gas law. That gives the non-linear equation
\begin{equation}
\begin{split}
     P_\text{compr}(q,p_\text{gas}^\text{in},p_\text{gas}^\text{out}) & =   c  \Bigl[(\frac{p_\text{gas}^\text{out}}{p_\text{gas}^\text{in}})^a - 1 \Bigr]  q
    \\
    & \quad a = \frac{k - 1}{k}
    \\
    & \quad c = \frac{\rho}{\eta_{is}} \frac{1}{k - 1}  Z  R  T_1
\end{split}
\end{equation}
where
$q$ is the gas flow rate,
$p_1$, $p_2$ is inlet/outlet pressure,
$R$ is the individual gas constant,
$\rho$ is gas density at standard conditions,
$\eta_{is}$ is the isentropic efficiency of the compressor,
$k$ is the specific heat capacity ratio,
$Z$ is gas compressibility,
and
$T_1$ is inlet temperature.

The linearised version of this power demand function is:
\begin{equation}
    P^\text{linear}_\text{compr}(q,p_\text{gas}^\text{in},p_\text{gas}^\text{out}) 
    =  c  \Bigl[
    a  (\frac{\hat{p}^\text{out}}{\hat{p}^\text{in}})^a \hat{q}
    (\frac{p^\text{out}}{\hat{p}^\text{out}} 
    -\frac{p^\text{in}}{\hat{p}^\text{in}})
    + ((\frac{\hat{p}^\text{out}}{\hat{p}^\text{in}})^a - 1
   ) q  \Bigr] 
\end{equation}
For an electric compressor we therefore get the following constraints
\begin{align}
    f^\text{gas}_\text{out} &=  f^\text{gas}_\text{in} 
    \\
    f^\text{in}_\text{el} &= P^\text{linear}_\text{compr}(f_\text{gas}^\text{out},p_\text{gas}^\text{in},p_\text{gas}^\text{out})
\end{align}
and for a gas-driven compressor:
\begin{align}
    f^\text{gas}_\text{out} &=  f^\text{gas}_\text{in} 
    -\frac{1}{c_\text{gas}} P^\text{linear}_\text{compr}(f_\text{gas}^\text{out},p_\text{gas}^\text{in},p_\text{gas}^\text{out})
\end{align}
where 
$c_\text{gas}$is the gas calorific value (combustible energy content).

\paragraph{Pumps}

Power consumption by oil or water pumps can be computed considering the work needed to move the liquid up a pressure gradient. 
For an incompressible liquid, the result is
\begin{equation}
    P_\text{pump}(q) = \frac{1}{\eta_\text{pump}} (p^\text{out}-p^\text{in}) q
    \approx \frac{1}{\eta_\text{pump}} (\hat{p}^\text{out}-\hat{p}^\text{in}) q
\end{equation}
where
$q$ is the flow rate,
$\eta_\text{pump}$ is the pump efficiency,
and we have assumed that pressure levels are close to nominal values $\hat{p}$
This gives the pump equations:
\begin{align}
    f^\text{out} &= f^\text{in}
    \\
    f^\text{in}_\text{el} &= P_\text{pump}(f^\text{in})
\end{align}

\paragraph{Gas turbine generators}
Gas turbine generators use gas to generate electricity and heat.
The equations expressing their relaionships are:
\begin{align}
	f &= f_\text{el}^\text{out}
	\label{eq:gt_f}
	\\
	\frac{f_\text{gas}^\text{in}}{f^\text{max}} c_\text{gas} &= A\frac{f_\text{el}^\text{out}}{f^\text{max}} + B(y_\text{on}+y_\text{prep})
	\label{eq:gt_fuel}
	\\
	f_\text{heat}^\text{out} &= (f_\text{gas}^\text{in} c_\text{gas} - f_\text{el}^\text{out})\eta_\text{heat}
	\label{eq:gt_flows}
\end{align}
where 
$f_\text{gas}^\text{in}$,
$f_\text{el}^\text{out}$ and
$f_\text{heat}^\text{out}$ are variables describing the flow in/out of the device,
$c_\text{gas}$ is the gas energy content (MJ/Sm3), 
$A$ and $B$ are dimensionless coefficients that specify the linear fuel consumption curve.


\paragraph{Heater}
For electric heaters (boilers or heat pumps), the equation relating electricity consumption and heat output is
\begin{align}
    f & = f_\text{el}^\text{in}
    \\
     f_\text{heat}^\text{out} &= \eta f_\text{el}^\text{in}
\end{align}
where
$\eta$ is the el-to-heat efficiency.

\paragraph{Sources}
These device only have an equation defining the flow variable $f$ used in the generic device constraints, depending on source type:
\begin{equation}
    f=f^\text{in}_\text{c}, 
    \quad
    c\in \{\text{oil},\text{gas},\text{water},
\text{el},\text{heat},\text{hydrogen}\}
\end{equation}

\paragraph{Sinks}
As for sources, sink device only have an equation defining the flow variable $f$ used in the generic device constraints, depending on which sink type:
\begin{equation}
    f=f^\text{in}_\text{c}, 
    \quad
    c\in \{\text{oil},\text{gas},\text{water},
\text{el},\text{heat},\text{hydrogen}\}
\end{equation}

\paragraph{Battery storage}
The constraints for battery storage are:
\begin{align}
    f &= f_\text{el}^\text{out}
    \\
     (\eta f_\text{el}^\text{in}- \frac{1}{\eta} f_\text{el}^\text{out}) \Delta t 
     &= E(t) - E(t-1)
     \\
     E^\text{min} \le & E \le E^\text{max}
     \\
     f^\text{out}_\text{el} &\le f^\text{max}
     \\
     f^\text{in}_\text{el} &\le -f^\text{min}
     \\
     p^\text{max} &= \min\{f^\text{max},\frac{E}{t_\text{res}}\}
\end{align}
where
$\Delta t$ is the timestep duration,
$\eta$ is the charge/discharge efficiency,
$E(t)$ is the storage energy content at timestep $t$,
and
$t_\text{res}$ is the minimum time the power must be available to count towards reserve.
The variable $p^\text{max}$ gives the maximum available power from the the storage, a quantity needed to compute the power reserve. It is limited both by the discharge capacity and by the energy in the storage.

The equation with the minimum value function is represented by the linear equations
\begin{align}
     f^\text{max} - M (1-y_\text{storage}) 
     &\le p^\text{max} 
     \le f^\text{max}
     \\
     \frac{E}{t_\text{res}} - M  y_\text{storage}
     &\le 
     p^\text{max} 
     \le \frac{E}{t_\text{res}}
\end{align}
where
$y_\text{storage}$ is a binary helper variable,
$M$ is a large (big-M) parameter (larger than $f^\text{max}$).
The binary variable $y$ will be 1 if  $\frac{E}{t_\text{res}}<f^\text{max}$, 
and 0 otherwise.

\paragraph{Hydrogen storage}
The hydrogen storage model is somewhat simpler than battery storage because the charging/discharging is represented by separate units (electrolyser and fuel cell). Its equations are
\begin{align}
    f &= f_\text{hydrogen}^\text{out}
    \\
     (f_\text{hydrogen}^\text{in}-f_\text{el}^\text{out}) \Delta t 
     &= E(t) - E(t-1)
     \\
     E^\text{min} & \le  E \le E^\text{max}
     \\
     \delta_E &\ge -(E-E_\text{target})
\end{align}
where $\delta_E$ is a non-negative variable representing the energy storage deviation from the target value $E_\text{target}$ at the end of the optimisation horizon.

The storage deviation constraints are associated with the storage depletion penalty in the objective function (Eq.~\ref{eq:objectivefunction}).

\paragraph{Electrolyser}
These devices produce hydrogen and heat from electricity:
\begin{align}
    f &= f^\text{el}_\text{in}
    \\
    f^\text{hydrogen}_\text{out} c_h &= f^\text{el}_\text{in} \eta_\text{h}
    \\
     f^\text{heat}_\text{out} &= f^\text{el}_\text{in}(1-\eta_\text{h})\eta_\text{heat}
\end{align}
where
$c_h$ is the hydrogen calorific value (energy content),
$\eta_\text{h}$ is the electrolyser efficiency,
and
$\eta_\text{heat}$ is the waste heat recovery efficiency.

\paragraph{Fuelcells}
These produce electricity and heat from hydrogen:
\begin{align}
    f &= f^\text{el}_\text{out}
    \\
    f^\text{el}_\text{out} &= f^\text{hydrogen}_\text{in} 
    c_h \eta_\text{cell}
    \\
    f^\text{heat}_\text{out} &= f^\text{hydrogen}_\text{in} 
    c_h (1-\eta_\text{cell})\eta_\text{heat}
\end{align}
where
$c_h$ is the hydrogen calorific value (energy content),
$\eta_\text{cell}$ is the fuel cell efficiency,
and
$\eta_\text{heat}$ is the waste heat recovery efficiency.

\subsection{Network constraints}
\label{sec:networkconstraints}
The following sections give the network-related constraints.

\subsubsection{Flow balance at each node terminal}

Edges \emph{to} the node $n$ are always connected to the \emph{in} terminal and edges \emph{from} the node are always connected to the \emph{out} terminal.
Therefore, for each  node $n$ and carrier $c$ the total flow into the "in" terminal and out of the "out" terminal are given as the contributions from the connected devices and edges as:
\begin{align}
    \text{in:} \qquad &
    F^\text{in}_{c} = 
        -\sum_{d\in\mathbb{D}_n} f_\text{d,c}^\text{in} 
        + \sum_{e\in\mathbb{E}^\text{to}_{c,n}} (q_e - q_{e,+}^\text{loss})
        - q^\text{term}_{c,n}
        = 0
    \\
    \text{out:} \qquad &
    F^\text{out}_{c} = 
        -\sum_{d\in\mathbb{D}_n} f_\text{d,c}^\text{out} 
        + \sum_{e\in\mathbb{E}^\text{from}_{c,n}} (q_e + q_{e,-}^\text{loss})
        - q^\text{term}_{c,n}
        = 0
\end{align}
where
$\mathbb{D}_n$ is the set of devices at node $n$,
$\mathbb{E}_{c,n}^\text{from}$ is the set of edges of carrier $c$ going from node $n$,
$\mathbb{E}_{c,n}^\text{to}$ is the set of edges of carrier $c$ going to node $n$.

The term $q^\text{term}_{c,n}$ is only present if there are no devices connecting both the in and the out terminal for a given carrier $c$ (and therefore have device constraints specifying the flow between them). It is used and is used to effectively merge the in/out terminals to a single terminal.

\subsubsection{Electrical losses}

To account for (electrical) losses, the edge flow, which in general can be positive or negative, is split in two always non-negative variables $q_+$and $q_-$, such that
\begin{align}
	q &= q_+ - q_-
\end{align}
Only one of these components will be non-zero at any time.

We define the flow components such that $q_+$ is the flow in positive direction out of the ``from" end and $q_-$ is the flow in negative direction out of the ``to" end.
%
Losses are calculated via a piece-wise linear function $F^\text{loss}$:
\begin{equation}
    q_+^\text{loss} = F^\text{loss}(q_+),
    \qquad
    q_-^\text{loss} = F^\text{loss}(q_-),
    \qquad
    q^\text{loss} =q_+^\text{loss} +q_-^\text{loss}.
\end{equation}

The general expressions for edge flow out of the ``from'' end ($q^\text{from}$) and into the ``to'' end ($q^\text{to}$) are:
\begin{align}
    q^\text{from} &= q_+-(q_--q_-^\text{loss})
    =q+q_-^\text{loss}
    \\
    q^\text{to} &=
    (q_+-q_+^\text{loss})-q_-
    =q-q_+^\text{loss}
\end{align}

\subsubsection{Fluid pressure deviation limit}
Limits on the deviation of pressure from the nominal value (given through input data) may be added for fluid flows. For example, if the gas pressure at the export point, or the water injection pump output pressure should have a fixed value, this may be enforced in this way.
The associated constraint is given as:
\begin{align}
    p_{n,c}^\text{nom}(1-\delta_{n,c}^\text{max}) 
    &\le p_{n,c} 
    \le  p_{n,c}^\text{nom}(1+\delta_{n,c}^\text{max}) 
\end{align}
where
$p_{n,c}$ is the pressure for carrier $c$ at node $n$,
$p_{n,c}^\text{nom}$ is nominal pressure,
and
$\delta_{n,c}^\text{max}$ is the maximum allowable relative pressure deviation from the nominal value (0--1).

\subsubsection{Edge flow limits}

Edges may be specified as bidirectional, which is generally the case for electric flow, or one-directional, which is generally the case for fluids:
\begin{align}
    -q^\text{max} &\le q \le q^\text{max} \qquad (\text{bidirectional})
    \\
    0 &\le q \le q^\text{max} \qquad (\text{one-directional})
\end{align}
where
$q$ is the edge flow,
$q^\text{max}$ is the maximum flow (in either direction).

\subsubsection{Edge flow equations}

The simplest way to represent edge flow is as a transport model, where the flow is constrained only by upper and lower bounds, and by energy/matter balance in the endpoint nodes. 

However, for extended networks with a meshed electricity grid or long-distance pipelines, it may be necessary to use more advanced models relating voltage angles and power flow for electric networks, or flow rate and pressure for pipeline fluid flows are possible. These are described below.

\paragraph{Electric power flow}

Power flow in an electricity grid is determined by the locations of power injection and ejection and the impedances of the alternative routes. The equations describing this are the non-linear AC power flow equations. A linearised version of these equations gives a good approximation if voltages are close to nominal values, phase angle differences are small and resistances are small compared to reactances. The linearised equations, in matrix form and expressed using dimensionless \emph{per unit} quantities, are
\begin{equation}
    \mathbf{Q} = \mathbf{DA\Theta},
\end{equation}
where
$\mathbf{Q}=[q_1,\dots,q_m]$ is a vector of edge power flow,
$\mathbf{D}=\text{diag}(-\frac{1}{x_1},\dots,-\frac{1}{x_m})$ is a diagonal matrix with elements given by the edge reactances,
$A$ is the $m\times n$-dimensional node--branch incidence matrix describing the network topology,
$\mathbf{\Theta}=[\theta_1,\dots,\theta_n]$ is the vector of nodal voltage angles,
and
$m$ is the number of edges and $n$ is the number of nodes.

In addition to the above equation, it is necessary to define a \emph{reference node} $n_\text{ref}$ for voltage angles such that $\theta_{n_\text{ref}}=0$.

\paragraph{Gas flow}

The empirical Weymouth equation is appropriate for describing pressure drop in gas pipelines with large diameter, at high pressure, and high flow rates \cite{menon2005}, and is used if pressure drop in gas pipelines is considered in the model.
The equation relates flow rate and pressure in and out:
\begin{equation}
 	 q = k\sqrt{\left( p_1^{2}-e^{s}p_2^{2} \right) }, 
\end{equation} 
where
$q$ is the edge fluid flow rate,
$p_1$ and $p_2$ are inlet and outlet pressure,
$\exp(s)$ is a constant factor that accounts for elevation differences (see below), 
and
$k$ is a pipeline property constant given as:
\begin{equation}
 	k=4.3328 \cdot 10^{-8}\frac{\mathrm{m}^{\mathrm{3}}}{\mathrm{\text{s MPa}}} \cdot  \frac{T_{b}}{P_{b}} \left[ GT_{f}L_{e}Z \right] ^{-\frac{1}{2}}D^{\frac{8}{3}}, 
\end{equation}
where
$D$ is pipe diameter (in mm),
$T_b$ is base temperature (ambient temperature) (in K),
$P_b$ is base pressure (local atmosphere) (in MPa),
$G$ is the gas gravity constant,
$T_f$ is the gas temperature, and
$Z$ is the compressibility factor.

The factor  $e^{s}$  and the equivalent length  $L_{e}$  together account for elevation differences. These are defined as:
\begin{equation} 
	s=0.0684 G\frac{z_{2}-z_{1}}{T_{f}Z}, 
\end{equation}
\begin{equation}
	 L_{e}=L\frac{e^{s}-1}{s}, 
\end{equation} 
where  $z$  is elevation (in m) and  $L$  is the segment length (km).

It is assumed here that the pressure remains close to the given nominal values  $\hat{p}_1$  and  $\hat{p}_2$ at the pipe endpoints. The non-linear Weymouth equation is then linearised around these points, \cite{Tomasgard2007}
\begin{equation}
 	 p_1 = \hat{p}_1 + \Delta p_1,
 	 \quad 
 	 p_2 = \hat{p}_2 + \Delta p_2,
\end{equation}
giving the linear equation:
\begin{equation}
	q =\frac{k}{
	\sqrt{ \left( \hat{p}_1^2-e^{s}\hat{p}_2^2 \right) }} 
	\left[ \hat{p}_1 p_1 - e^{s} \hat{p}_2 p_2 \right] ,
\end{equation}
that relates flow rate $q$ and pressure at start $p_1$ and endpoint $p_2$.

\paragraph{Liquid flow}
The pressure loss in an incompressible fluid (liquid) pipeline due to friction is described by the empirical Darcy-Weisbach equation \cite{menon2005}: 
\begin{equation}
 p_2-p_1
 	=- \rho g ( z_2-z_1) -f_D\frac{ \rho L}{D}\frac{V^{2}}{2}, 
 \end{equation}
where 
$p_1$ and $p_2$ are inlet and outlet pressure,
$V$ is average fluid velocity,
$z$ is elevation,
$\rho$ is fluid density,
$g=9.98~\text{m}/\text{s}^2$ is the gravity constant,
$L$ is the pipe length,
$D$ is the pipe diameter, and
$f_D$ is the Darcy friction factor.
 
Since the velocity  $V$ is related to the volumetric flow rate $q$, this can be written as
\begin{equation}
	p_2-p_1 = - \rho g ( z_2-z_1) -\frac{8f_D \rho L q^{2}}{ \pi ^{2}D^{5}}. 
\end{equation}
The Darcy friction factor $f_D$ is determined by pipe properties and flow regime. It can be computed in different ways, but in our model is assumed given as user input.

Linearising this equation around an operating point given by the nominal pressure values $\hat{p}_1,\hat{p}_2$ and  flow rate $\hat{q}$ (derived from the equation above), we get a linear equation that relates flow rate and in/out pressure for liquid pipeline flow:
\begin{align}
    p_2 - p_1 =& - (q-\hat{q}) \frac{2X}{k} 
        +(\hat{p}_2 - \hat{p}_1)
    \\
    &X = \sqrt{(\hat{p}_1 - \hat{p}_2) - \rho  g (z_2-z_1)}
    \\
    &k = \sqrt{\frac{\pi^2 D^5}{8 f_D \rho  L}}
    \\
    &\hat{q} = k X
\end{align}

Note that for this linear approximation to be good, the flow rate needs to be close to the nominal value. If conditions change, the coefficients should be updated.

\subsection{Global constraints}
\label{sec:globalconstraints}

In addition to constraints associated with devices and the networks, the optimisation problem may includes global constraints as explained in the following.

\paragraph{Reserve margin limit}
Power reserve is the online available power capacity that is not used. Both generators and loads may contribute reserve. The reserve requirement constraint is:
\begin{equation}
	\sum_{d\in\mathbb{D}} ( f_\text{d,el}^\text{max} x_d^\text{res} 
	- f_\text{d,el}^\text{out} 
	+ f_\text{d,el}^\text{in} x_d^\text{L,res})  \ge R^\text{L,min}
\end{equation}
where
$\mathbb{D}$ is the set of devices,
$f_\text{d,el}^\text{max}$ is the maximum available electric power output for device d,
$x_d^\text{res}$ is a \emph{reserve factor} parameter (0--1) that determines how much of the unused capacity should count towards the reserve,
$x_d^\text{L,res}$ is a parameter (0--1) that determines how much of the load can be reduced,
$R^\text{min}$ is the minimum reserve limit (MW).

For a generator, the reserve factor is typically 1, but may be less if for example there is uncertainty in the estimation of available capacity, as is the case for wind power.
For loads, the reserve factor $x_d^\text{L,res}$ is 0 if no load shedding is allowed.

\paragraph{Emission rate limit}
The \COO emission rate is the emission per time. If a hard limit on emissions should be enforced, an emission rate limit may be specified through the equation:
\begin{equation}
	\sum_{d\in \mathbb{D}_{gas}} c_\text{\COO} (f_\text{d,gas}^\text{in}-f_\text{d,gas}^\text{out} ) \le e_\text{\COO}^\text{max}
\end{equation}
where
$\mathbb{D}_{gt}$ is the set of gas-combusting devices (gas turbine generators and gas-driven compressors),
$c_\text{\COO}$ is gas \COO content (kg/Sm3),
and
$ e_\text{\COO}^\text{max}$ is the \COO emission rate limit (kg/s).

\bibliography{bibliography}

\end{document}